\documentclass[12pt]{article}

\usepackage{tikz}
\usepackage{amsmath}
\usepackage{comment}
\usepackage{amsthm}
\usepackage{amsfonts}
\usepackage{amssymb}
\usepackage{graphicx}
\usepackage{caption}
\usepackage{subcaption}
\usepackage{mathptmx}
\usepackage{stfloats} 
\usepackage{IEEEtrantools}
\usepackage{microtype}
\usepackage{listings}
\usepackage{tabularx}
\usepackage{epstopdf}
\usepackage{xcolor}
\usepackage{nomencl}
\usepackage{pifont}
\usepackage{float}
\usepackage{ulem}
\restylefloat{figure}
\usepackage[numbers,sort&compress]{natbib}
\usepackage{enumerate} 
\usepackage{fancyhdr}
\usepackage{lineno,hyperref}
 \usepackage{xcolor, soul}
\sethlcolor{green}

\newtheorem{theorem}{Theorem}

\newtheorem{definition}{Definition}

\newtheorem{proposition}[theorem]{Proposition}

\title{A sub-Riemannian model of the motor cortex with Wasserstein distance. }
\date{}\author{Jawad Ali, Giovanna Citti, Alessandro Sarti}
\begin{document}

\maketitle



\begin{abstract}
This study aims to better understand the functional geometry of the motor cortex, starting from different sources of experimental evidence. Recent studies have proved that cells of the primary motor cortex (M1) are sensitive to short hand trajectories called fragments.  Here, we propose a sub-Riemannian higher-dimensional geometry accounting for geometric and kinematic properties. Due to the constraints of the geometry,  horizontal curves naturally satisfy a relation between geometric and kinematic properties experimentally observed. In the space of trajectories, we also apply a clustering algorithm based on the  Wasserstein distance: we obtain a grouping which nicely fits the observed experimental data much more efficiently than the Sobolev distance.
\end{abstract}

\textbf{Keywords :} Primary motor cortex - Movement decomposition - Neurogeometry -Sub-Riemannian geometry- Wasserstein Distance

\tableofcontents

\section{Introduction}
This study aims to develop a model of the functional architecture of motor cortical cells in the primary motor cortex that control arm reaching movement in the brain, in order to study the geometry of movement and its cortical coding.
A first problem regarding the geometry of the movement is that there is a relation between the speed of movement and the curvature of the trajectory: the effector moves more slowly in the more curved parts of the trajectory. The first results in this topic are due to \cite{binet1893vitesse}, who proved that the speed $v$ of movement is proportional to the radius $R$ of the circle osculating the trajectory. More accurate measurements established the so-called 2/3 law between radius and speed (see \cite{bennequin2009movement}).

Flash and Handzel \cite{flash2007affine,handzel1999geometric}, Pollick and Shapiro \cite{pollick1997constant,shapiro1994motion} have remarked that this law \cite{bennequin2009movement} can be interpreted in terms of affine geometry: affine transformations are obtained by composing translations, rotations, stretching and dilatations, and a geometric property is affine invariant if it is preserved under affine transformations. In \cite{bennequin2017several} several types of transformations are considered: Euclidean, affine and equi-affine transformations which preserve both parallelism and area. There is not yet a complete explanation of this phenomenon, even though it is generally accepted that it can be related to the functional geometry of the brain.

The first experimental results regarding the functional architecture of motor cortex were obtained by A. Georgopoulos (see \cite{georgopoulos1984static, kettner1988primate} and \cite{georgopoulos1982relations, schwartz1988primate}): he proved that this area of the brain is sensitive to kinematic features such as the position $x, y$ of the hand, the direction $\theta$ of movement. Indeed, according to his measurements, the neural cell response is maximal when hand position, orientation, or speed coincides with the one characteristic of the cell.

\begin{figure}[htbp]
\centering
\includegraphics[scale= 0.42]{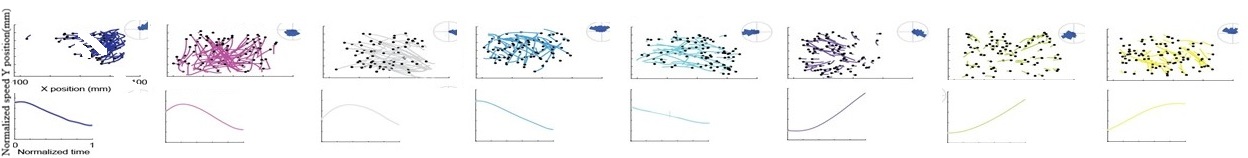}
\caption{The results of clusterization made by \cite{kadmon2019movement} group fragments in neural states: each one is a set of fragments with homogeneous orientation and increasing or decreasing acceleration phase. Within each column, we see the $(x, y)$ section of a set of fragments (above) and the corresponding profile in the  $(t, v)$ plane (below).}
\label{clusterini}
\end{figure}

After that, it was proved that the activity of neurons in the primary motor cortex is sensitive to velocity and acceleration (see for example \cite{kettner1988primate}, \cite{moran1999motor}), as well as to joint angles (see \cite{ajemian2001model}, \cite{teka2017motor}).
Later on, Hatsopoulos (\cite{Encoding, reimer2009problem}) proved that the tuning for movement parameters are not static, but M1 encodes shorthand trajectories called fragments. This study is compatible with the general approach of Graziano, who postulated that the motor cortex is organized into action maps (see \cite{graziano2002cortical, graziano2007mapping}). The problem was further investigated by H.N. Kadmon \cite{kadmon2019movement}, who recorded neural activity with a 100-electrode array, processed the recording with a grouping algorithm and identified sequences of coherent behaviours, called neural states. The result of clusterization is presented in Figure \ref{clusterini}, where each state contains a family of elementary trajectories, characterized by a specific orientation, and by an acceleration or deceleration phase.

Mathematical models of this brain area were proposed by various authors, with the specific scope to understand brain organization and its geometry. A trajectory's smoothness criterion was initially defined as the minimization of the integrated squared rate of movement jerk, which is the derivative of the acceleration \cite{flash2001computational}.
Other models of fragments are proposed by \cite{flash2007affine, maoz2014spatial, polyakov2017affine}. However, the problem of the geometry of the motor cortex, the coding of signals, and in particular properties of fragments was still an open problem. The first sub-Riemannian model of motion was introduced in Jean's book \cite{jean2014control, jean2010optimal}. Another model was introduced in \cite{mazzetti2023functional}. In this paper, the authors model the space of features as the set
\begin{equation}\label{M}(x, y,t, \theta, v, a) \in \mathcal{M}:=\mathbb{R}_{(x,y)}^2  \times \mathbb{R}_{t} \times S_\theta^{1} \times \mathbb{R}_{v} \times \mathbb{R}_{a},\end{equation} 
where the features are the same as before: $x$, $y$ are the position of the hand,
 $\theta$ the direction of movement, $v$ and $a$ the velocity and acceleration. 
Then, they induced in this manifold a sub-Riemannian structure (see \eqref{eq11} below), and proposed to model fragments as sub-Riemannian  horizontal  curves 
(see \eqref{fragments} for the definition). In addition, in \cite{mazzetti2024sub},  a first grouping algorithm in this space of 'ideal' fragments was proposed, able to recover a clusterization similar to Figure \ref{clusterini} on artificial data, all defined on the same time interval, with  a Sobolev-type distance. 

In this paper, we provide a new sub-Riemannian model of the motor cortex able to give an answer to the 
problems described above. The first one is the relation between the radius of the osculating circle of a trajectory and the speed of movement: the answer to the first problem relies on the properties of the sub-Riemannian geometry itself. Indeed, this is defined in terms of the natural differential constraints which relate the variables of the space: position and orientation on one hand, time, position, velocity and acceleration on the other.
The large majority of the paper is devoted to facing the problem of clusterization of fragments in neural states. We are inspired by \cite{mazzetti2023functional}, and we will look for a distance in the space of fragments, but we aim to remove the assumption that all trajectories are defined on the same unitary interval.
 Indeed, fragments are parametrized by time, which represents an absolute and external variable, which cannot be reparametrized or warped without altering the meaning of the process. For this reason, we propose here to use the Wasserstein distance, and show that it provides reliable results even on real data and fragments, whose initial and final instants of time are not a priori fixed.
Our grouping correctly fits the cluster experimentally obtained by working with the cortical activity. This will confirm that the geometry we propose is a good model of the cortical connectivity of this area. The Wasserstein distance is introduced here for the first time in modelling functional architecture of the cortex, but we believe that it is the natural instrument to describe the cortical connectivity, due to its probabilistic nature, and the possibility to define it on progressively more complex and abstract cortical areas.

The structure of the paper is as follows. In Section \ref{stateofthe art}, we present the state of the art of modelling in this area, and the mathematical instruments needed to formalize our model: the sub-Riemannian geometry and the Wasserstein distance. In Section \ref{thegeometryofthemotorcortex}, we show that the model of fragments correctly codes the relation between geometry and kinematics. Then we formulate our model of neural states in terms of the Wasserstein distance. In Section \ref{results}, we test the algorithm on artificial and real data \cite{kadmon2019movement}.

\section{The state of the art}\label{stateofthe art}

\subsection{A sub-Riemannian geometry of the cortex and  fragments}

 Following the first models of \cite{georgopoulos1981spatial}, we can assume that cells in the motor cortex are selective to the  following features:  
hand position in two dimensions $(x, y)$,  instant of time $t$,  direction $\theta$ of movement, its    velocity $v$ and acceleration $a$. Consequently, the  set of features in the motor cortex can be modeled by the following variables  
\begin{equation}\label{M}(x, y,t, \theta, v,a) \in \mathcal{M}:=\mathbb{R}_{(x,y)}^2  \times \mathbb{R}^{+}_{t} \times S_\theta^{1} \times \mathbb{R}_{v} \times \mathbb{R}_{a},\end{equation} 

Let us note in particular that these variables contain both geometric (such as orientation) and kinematic properties (velocity and acceleration) of the trajectory. By using the differential constraints between the variables of this space, the authors of \cite{mazzetti2023functional}  introduced the following vector fields
\begin{equation}\label{eq11}
X_1 = v\cos\theta \frac{\partial}{\partial x} + v\sin\theta \frac{\partial}{\partial y} + a \frac{\partial}{\partial v} + \frac{\partial}{\partial t}, \quad X_2 = \frac{\partial}{\partial \theta}, \quad X_3 = \frac{\partial}{\partial a}.
\end{equation}
and defined in $\mathcal{M}$ a  sub-Riemannian structure
$(\mathcal{M}, \Delta, g)$
, where $\Delta$ is the horizontal tangent space spanned by $X_1, X_2, X_3$, and $g$ is the metric which makes them orthonormal. This means that all the differential objects of the space have to be expressed in terms of these vector fields. In particular, horizontal curves are integral curves of vector fields, 
$$\gamma' (s) = \alpha_1X_1 + \alpha_2 X_2+ \alpha_3 X_3.
$$
In addition, a particular class of these curves, called admissible curves, has been proposed as a good model of fragments.  
The space of fragments is then modelled as: 
\begin{align}\label{fragments}
\mathcal{F}=\Big\{\gamma:
[\alpha,\beta] \to \mathcal{M}: \gamma' (s) = \alpha_1X_1 + \alpha_2 X_2+ \alpha_3 X_3:   \quad \quad \quad\quad \quad \quad \\
\quad \quad \quad\quad \quad \quad  \alpha_1, \alpha_2, j \in R, \;\; \alpha_3(t) =  j\Big(t- \frac{\alpha + \beta}{2}\Big)\;\Big\}.\nonumber
\end{align} 
We refer to \cite{mazzetti2023functional} where the expression of $\alpha_3$ was proposed in order to fit the typical shape of acceleration, measured in \cite{Hatsopoulos}. In \cite{mazzetti2023functional}, a change of variable was introduced, and all the curves were parametrized by a new parameter $s,$ different from the time parameter $t$,  in such a way as to be defined on the same interval $[0,1]$. 
Consequently, every curve will be represented as $$\gamma(s) = (x(s), y(s),  \theta(s), t(s), v(s), a(s)).$$
The feature space $\mathcal{M}$ has dimension 6. The Lie 
algebra generated by $X_1, X_2, X_3$ and their commutators 
spans all of the tangent space to  $M$, so that the H\"ormander condition 
is satisfied and the Carnot-Carath\'eodory distance 
$d_{\mathcal{M}}$ is well defined. Due to the works of Nagel et al.\cite{nagel1985balls} and 
Montgomery\cite{montgomery2002tour}, the Carnot-Carath\'eodory distance can be 
locally approximated by a homogeneous distance:
\begin{equation}\label{subD}
 d_{\mathcal{M}}(\eta_0,\eta) \simeq 
\bigl(c_1|e_1|^6 + c_2|e_2|^6 + c_3|e_3|^6 
+ c_4|e_4|^3 + c_5|e_5|^3 + c_6|e_6|^2\bigr)^{\frac{1}{6}},   
\end{equation}

where $c_i$ are non-negative constant coefficients and the 
number 6 is the dimension of $\mathcal{M}$.We will experimentally fix the values $c_i$ in section \ref{clusreal}. In \cite{mazzetti2023functional}, a spectral clustering algorithm 
is applied in this space to group kinematic features into 
fragments, by means of a kernel 
\begin{equation}\label{KM}K_{\mathcal{M}} = e^{-\frac{d_{\mathcal{M}}}{\sigma}}.
\end{equation}
for a suitable value of $\sigma$

\subsubsection{A Sobolev distance in the space of fragments}

The main idea developed in \cite{mazzetti2024sub} is to apply a geometric clustering algorithm in the space of features to obtain fragments, and then to apply the same algorithm, with a different distance to find neural states. Since the clustering in neural states proposed in \cite{kadmon2019movement} appears to be independent of $x$ and $y$, the authors of \cite{mazzetti2024sub}  introduced a subset 
${\mathcal{M}}_1$ of ${\mathcal{M}}$ independent of the variables $(x,y)$ 
$${\mathcal{M}}_1 = \{(x, y, \theta, t, v, a): (x,y)=(0,0)\} $$
and considered  on  ${\mathcal{M}}_1$
the distance $d_{\mathcal{M}_1}$ 
induced by the immersion in $\mathcal{M}$. The distance $d_{\mathcal{M}_1}$ can be extended to a pseudo-distance 
in the space $\mathcal{M}$ by setting 
$$d_{\mathcal{M}_1}((x, y, \theta, t, v, a), (x', y', \theta', t, v', a'))= 
d_{\mathcal{M}_1}((0, 0, \theta, t, v, a), (0, 0, \theta', t, v', a'))
$$
Let us explicitly note that it naturally defines a pseudo-distance in the space $\mathcal{F}$  as follows. First of all, it is necessary to introduce a reparametrization of the curves, and assume that they are all defined on the same interval $[0,1],$ then the following Sobolev-type distance can be introduced. 
 \begin{definition}
If $\gamma_1, \gamma_2 \in \mathcal{F}$, then we can call 
\begin{equation}\label{sobd}
  d_\mathcal{F}(\gamma_1, \gamma_2) = \int_0^1 ||\gamma'_1(s)-\gamma_2'(s)||_{\mathcal{M}_1} ds +
d_{\mathcal{M}_1}(\gamma_1(1),\gamma_2(1)) .   
\end{equation}

 \end{definition}

\bigskip

Then the authors introduced  a kernel, in terms of the distance, which models the propagation of the signal along connectivity in the space of fragments,  
\begin{equation}\label{connectivityF}
K_{\mathcal{F}}(\gamma_i, \gamma_j) = e^{-\frac{d_{\mathcal{F}}(\gamma_i, \gamma_j)^2}{\sigma}}.
\end{equation}

Using this kernel and a geometric spectral clustering algorithm on a set of artificially generated curves, the authors recovered a 
 classification compatible with the one in \cite{kadmon2019movement}: 

\begin{figure}[H]
\centering
\includegraphics[width=7 cm]{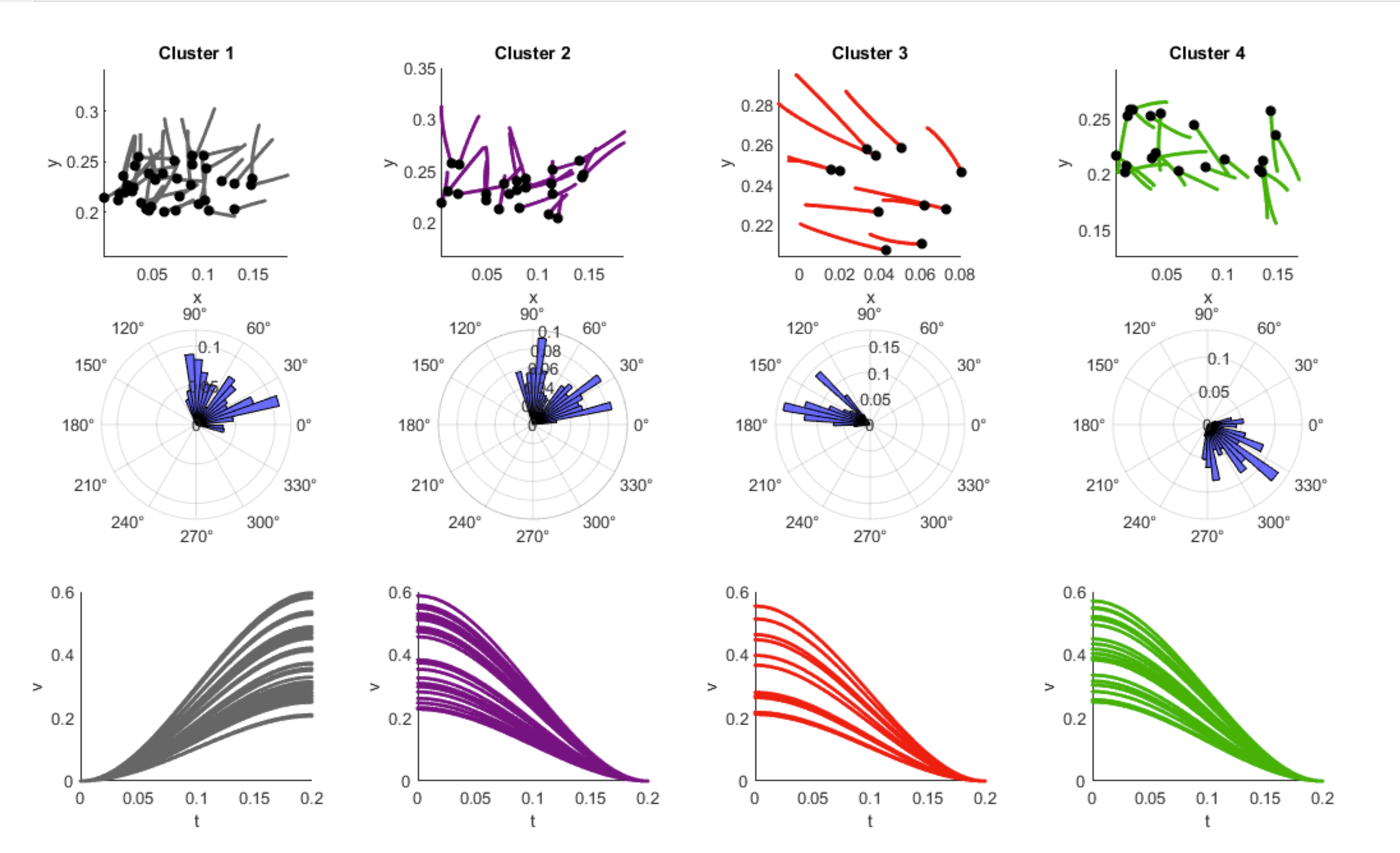}\hspace{ - .4 cm}
\includegraphics[width=7 cm]{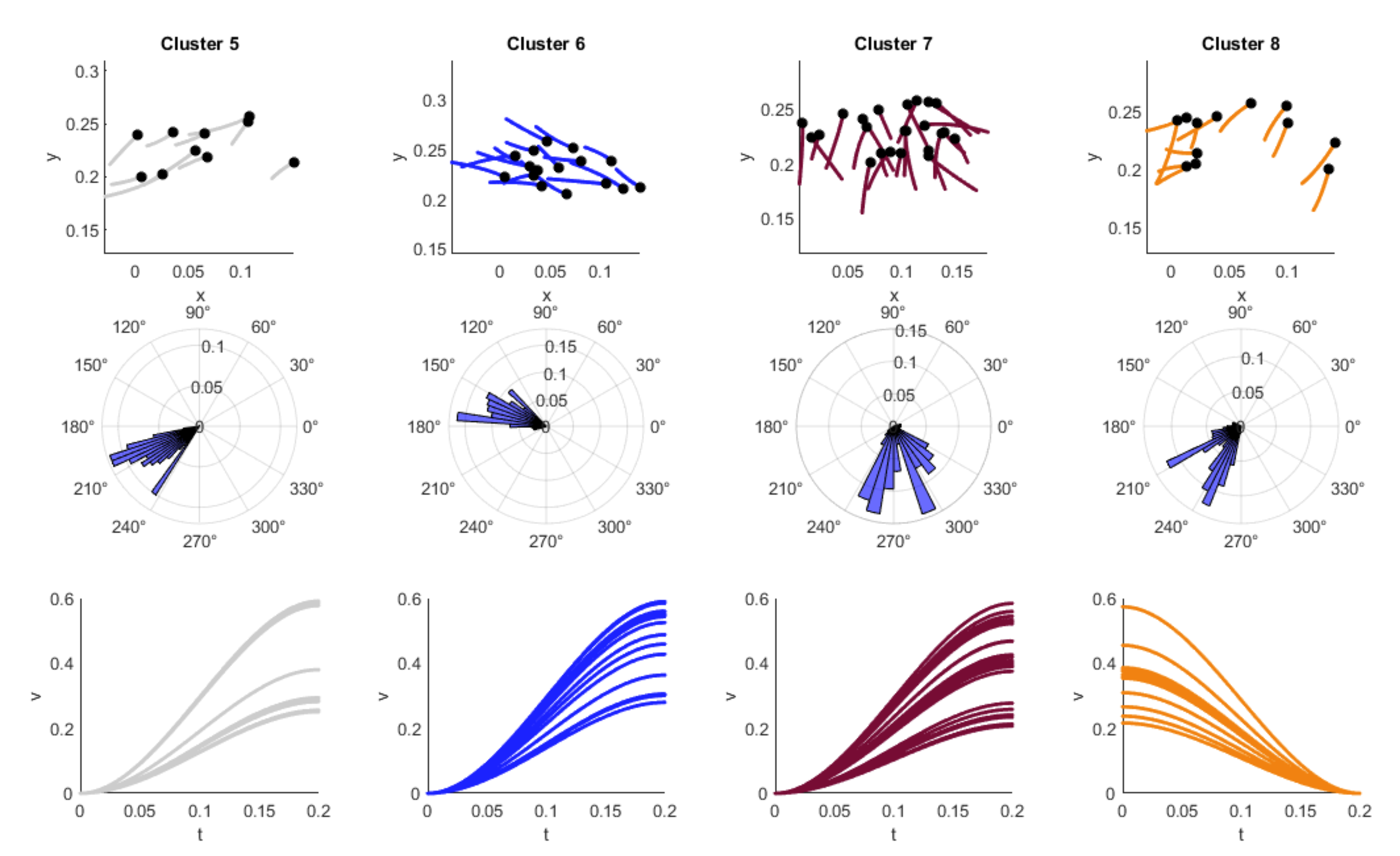}
\caption{Visualization of the grouping of fragments in neural states obtained in \cite{mazzetti2024sub}. }
\label{002}
\end{figure} 

As it is clear from Figure \ref{002} that the algorithm performs well on artificially generated curves in reproducing the classification obtained in \cite{kadmon2019movement}. 
Consequently, we can deduce that the  Sobolev distance can describe connectivity in spaces of functions. Under the assumption that all trajectories are generated on the same time interval, the Sobolev distance could be reliably used. However, the algorithm simply does not work on real data, defined on fragments which are defined on different time intervals  (see Figure \ref{11133}). 
To compare two trajectories with varying speed, we required a distance metric that: (i) handles irregular sampling naturally, 
(ii) preserves biologically meaningful speed information, and (iii) allows 
temporal alignment without duration normalization.

We evaluated Wasserstein distance (detailed in Results) and this distance treats trajectories 
as distributions, accommodating variable lengths without preprocessing, while 
Sobolev requires uniform time grids necessitating interpolation artifacts. Normalizing these temporal differences 
would eliminate precisely the speed-dependent neural activity patterns we seek 
to distinguish. Wasserstein distance avoids these issues by comparing velocity 
distributions directly, preserving both the natural timing and magnitude of 
movements without requiring preprocessing.
Based on theoretical considerations \cite{churchland2012neural} and empirical 
validation (see Results), we employed the Wasserstein distance for all analyses.

\subsection{The Wasserstein distance}

In this paper, we propose to use the Wasserstein distance instead of the Sobolev one. Indeed, it is stable under translation on the domain. 
In addition, since this has a probabilistic meaning, it is reasonable to expect that it is more robust to noise. 
To this end, 
we provide here the Kantorovich definition of the distance

\begin{definition}
  Given a measure \( \pi \) on \( X \times Y \), the marginals \( \mu \) and \( \nu \) are defined as:
\[
\mu(A) = \pi(A \times Y) \quad \forall A \subset X,
\]
\[
\nu(B) = \pi(X \times B) \quad \forall B \subset Y.
\]  
\end{definition}

\begin{definition}
    Let \( \Pi(\mu, \nu) \) denote all the couplings between $\mu$ and $\nu$ that have marginals \( \mu \) and \( \nu \). Let \( p \geq 1, \) the  Wasserstein $2$-distance is defined as:
    \[
    W_2(\mu, \nu) = \left( \inf_{{\Pi}_{0} \in \Pi(\mu, \nu)} \int \|x - y\|^2 \, d {\Pi}_{0} \right)^{1/2},
    \]
    
  The minimizer \( \Pi^* \) exists, and it is called the optimal transport plan or the optimal coupling (see for example \cite{villani2008optimal}).

\end{definition}

\subsubsection{Distance for signed and vector valued probability measures}

Wasserstein distance can also be naturally extended to non-negative measures as follows: 
\begin{definition} \cite{piccoli2019wasserstein}
    Let $\mu = \mu^+ - \mu^-
$, $\nu = \nu^+ - \nu^-
$ be two signed measures in $\mathcal{M}_s(\mathbb{R}^d)$. Then we define
\[
W_{2}(\mu, \nu) = W_{2}(\mu^+ + \nu^-, \mu^- + \nu^+),
\]

where $\mu^+$, $\mu^-$, $\nu^+$, and $\nu^-$ are any measures in $\mathcal{M}(\mathbb{R}^d)$ such that $\mu = \mu^+ - \mu^-$ and $\nu = \nu^+ - \nu^-$.
\end{definition}

The definition of Wasserstein distance for vector-valued measures is:
\begin{definition}\cite{zhu2022vwcluster}
Let $\Vec{\mu}= (\mu_1, \mu_2, \mu_3,..,\mu_n)$ and $\Vec{\nu}= (\nu_1, \nu_2, \nu_3,..,\nu_n)$  be vector-valued probability densities. The vector-valued Wasserstein distance between these densities is defined as:
\[
W_2(\Vec{\mu}, \Vec{\nu}) =\left(  \sum_{i=1}^{n} W_2(\mu_i,\nu_i)^2\right)^{1/2}.
\]    
\end{definition}

\subsubsection{The one-dimensional case}
When dealing with one-dimensional distributions, the computation of the Wasserstein distance becomes more straightforward. 
Indeed,
\begin{definition}
    The \textit{cumulative distribution function (CDF)} of a probability distribution \( \mu \) is a function \( F_\mu: \mathbb{R} \to [\alpha, \beta] \) defined by:
    \[
    F_\mu(x) = \mu(X \leq x),
    \]
    where \( X \) is a random variable with distribution \( \mu \).  
\end{definition}

For distributions \( \mu \) and \( \nu \) on the real line, the \( 2 \)-Wasserstein distance \( W_2 \) \cite{kolouri2017optimal} has a closed form:

\begin{proposition}

\begin{equation}\label{Wass-d}
   W_2(\mu, \nu) = \left( \int_\alpha^\beta \left| F_\mu^{-1}(z) - F_\nu^{-1}(z) \right|^2 \, dz \right)^{1/2}, 
\end{equation}
where \( F_\mu^{-1} \) and \( F_\nu^{-1} \) are the inverses of the CDFs for the probability measures \( \mu \) and \( \nu \).
    
\end{proposition}

\section{The geometry of the motor cortex}\label{thegeometryofthemotorcortex}

The scope of this work is to show that the sub-Riemannian geometry is able to clarify two main aspects of the geometry of the motor cortex. The first aspect is the relation between kinematic and geometric variables of locomotion, observed in literature, which in our setting will be deduced by the differential relation which defines the sub-Riemannian space. The second is the classification of motor primitives into neural states. We will show that the  Wasserstein distance in the space of curves, is able to recover clustering of primitives comparable with the experimentally observed ones. 

\subsection{Geometry of spaces of features and fragments}

The space $\mathcal{F}$ of fragments has been introduced in \ref{fragments}. In \cite{mazzetti2023direction}, up to a parametrization, it was assumed that all curves were defined on the same interval $[0,1].$ 
In this paper, we would like to present a reliable cortically inspired model. The  time variable is an external variable, and we cannot assume that the brain can re-parametrize it. Hence we will parametrize all curves via the time parameter $t$ and we will allow every curve to be defined on a different interval $[\alpha, \beta].$

\subsubsection{Kinematic and geometric variables}

Following the approach of \cite{mazzetti2023functional}, we will consider the space of features defined in \eqref{M}. As recalled above, admissible curves of the space, are solutions of the  ODE \eqref{fragments}; 
$$\gamma' = X_1 + \alpha_2 X_2 + \alpha_3 X_3,$$
where the vector fields $X_i$ are defined in \ref{eq11}
and 
$\alpha_3$ is a continuous function. 
 In particular, a solution satisfies: 
 
$$x' = v \cos(\theta),  y' = v \sin(\theta).$$
Since 
$$x'' = v' \cos(\theta) - v \sin(\theta)\theta', \quad y'' = v' \sin(\theta) +  v \sin(\theta)\theta',$$
then the curvature can be computed as
\begin{equation}\label{curv}k = \frac{x''y'- x' y''}{((x')^2 + (y')^2)^{3/2}} = \frac{\alpha_2}{v}\end{equation}
This relation is quite interesting since it implies a relation between the  radius $R$ of the osculating circle of a  trajectory and  the speed of the trajectory: 
\begin{equation} \label{RV}R= \frac{1}{\alpha_2} v
\end{equation}

 This is not an assumption; it is an algebraic consequence of the horizontal constraint. The sub-Riemannian geometry is the optimal choice precisely because the
speed–curvature relation emerges automatically from the differential constraints. The interest of this relation is that a relation has been experimentally observed between speed and radius as recalled in the Introduction. 
Simply speaking, circles with a small radius are traveled more slowly than those with a large radius. 

 Here, we claim that this relation is already encoded in our choice of variables.  We can also recover the relation $R= C v^3$ by simply choosing  $\alpha_2 = v^{-2}$, which corresponds to equi-affine parameterization 
\cite{pollick1997constant}. 
Let us explicitly note that the condition that $R=\frac{1}{\alpha_2}v$
is a consequence that we chose a planar retina. If we take into account the geometry of the retina, the relation $\theta'=k$ is replaced by a more sophisticated relation, which induces a different relation between $R$ and $v$.

\subsubsection{Pseudo-distances induced by restriction of subspaces}

In  \cite{mazzetti2024sub}, the authors introduced a pseudo-distance on the space, reducing to a subspace independent of the variables $x, y$. Precisely, they  introduced a sub-manifold $\mathcal{M}_1$ of  the feature space, defined as $$\mathcal{M}_1 = \{(t, \theta, v,a)\in  \mathbb{R}^+_t\times  S^1_{\theta} \times \mathbb{R}^{2}_{\left(v,a\right)}\}.
$$ 
There is a natural differential relation between the variables of the space, expressed by the-form $$\omega = adt - dv =0. $$
A flat Euclidean metric on $(x,y,\theta,v,a,t)$ treats all variables as independent, encoding no coupling between $\theta$ and $(x, y)$-motion or between $v$ and $a$. In a Riemannian setting, all tangent directions are equally accessible. On the other side the variables $x,y,\theta$ from one side, $t, v, a$ from the other are related by differential constraints. In case of the variables $t,v,a$ the non-holonomic constraint $\omega = a\,dt - dv = 0$  defines the acceleration in terms of the velocity.  This constraint reduces the dimension of the admissible tangent space, and defines a horizontal distribution $\Delta$ as the kernel of $\omega$. In this way the Sub-Riemannian tangent space is naturally introduced.

For this reason, we choose the generators of the kernel of $\omega$ as a basis of the horizontal tangent plane at every point: 

\begin{equation}
    \hat{X}_{1} = a\frac{\partial}{\partial{v}}+ \frac{\partial}{\partial{t}}, \quad 
    \hat{X}_{2} = \frac{\partial}{\partial{\theta}}, \quad 
    \hat{X}_{3} = \frac{\partial}{\partial{a}}.
\end{equation}
We will also call $$\hat{X}_{4} = [\hat{X}_{3}, \hat{X}_{1}] = \frac{\partial}{\partial{v}}.$$

The Hörmander condition is satisfied, since the Lie brackets of these vector fields generate the entire tangent space at every point. Hence the CC-distance can be estimated in terms of the ball box distance. Specifically, if we fix the coefficients $c_1, ... , c_4$, we can define the  norm of a vector $e = \sum_{i=1}^4 e_i \hat X_i$ as follows
\begin{equation}\label{distanza_periodic}
||e||_{\mathcal{M}_1}=\left(c_1\left|e_{1}\right|^{2} + c_2\left|e_{2}\right|^{2}+ c_3\left|e_{3}\right|^{2}+ c_4\left|e_{4}\right| \right)^{\frac{1}{2}}.
\end{equation}

The parameters $c_1, ... , c_4$ in \eqref{distanza_periodic} can be arbitrary. In the sequel, they will be tuned in such a way to obtain the experimentally observed clusterization.

A distance in the space $\mathcal{M}_1$
will be defined as following in terms of the canonical coordinates of any point
\begin{definition}
Let $\hat \eta_0\in\mathcal{M}_1$ be fixed. We define the canonical coordinates  $ {\hat \Phi}_{\hat\eta_0}(\hat \eta) $   of $\hat \eta$ around a fixed point $\hat\eta_0$, as the coefficients $e = (e_1, \cdots, e_4) $ such that 
\begin{equation}
\hat\eta= \exp\left(\sum_{i=1}^4 e_i \hat X_i\right)\left(\hat \eta_0\right).
\end{equation}

The distance in ${\mathcal{M}_1}$ naturally induces a pseudo-distance on ${\mathcal{M}_1}$
Consequently, we define 
$${\hat d}_{\mathcal{M}_1}(\hat\eta, \hat\eta_0) = ||{\hat \Phi}_{\hat\eta_0}(\hat \eta)||_{\mathcal{M}_1}.$$ 
\end{definition}

\subsubsection{Wasserstein distances in the space of fragments}
Here, we associate to each curve $\gamma = (x, y, \theta, v, t, a)$ a probability measure. We will be interested in both real and vector-valued densities. 
We define a real-valued measure as following
\begin{equation}\label{cifa}
\mu_{0}(\gamma)=\frac{||\gamma'(t)||}{\int_\alpha^\beta ||\gamma'(t)|| dt}.
\end{equation}
The definition of $\mu_0(\gamma)$ is biologically motivated: neuronal 
activity in M1 is expressed as a firing rate, representing the 
probability that a given set of kinematic features is attained along 
the trajectory \cite{georgopoulos1982relations}. 
The measure $\mu_0(\gamma)$ assigns higher mass to regions of 
$\mathcal{M}$ more frequently visited by $\gamma$, thus encoding 
exactly this probability structure.

We will also associate to any curve  $\gamma= (x,y,\theta, v,t,a)  $   a vector-valued signed  density 
$\mu =(\mu_1, \mu_2, \mu_3)$. 
We start with a curve and we consider only three components $$\big(\cos(\theta(s)  )  , \sin(\theta(s)  )  , a(s)  \big)  .$$

Then we define 

\begin{equation}\label{density}
 \mu(\gamma(s)  )   =  \frac{\Big( \sin(\theta(s)  )  ,\cos(\theta(s)  )  , a(s)   \Big)    }{\int_\alpha^\beta \sqrt{1+a^2(s)  } ds},   
\end{equation}

If we have two curves
$\gamma_1$ 
and $\gamma_2,$ the corresponding distances will be 
\begin{equation*}
d_{0, \mathcal{F}}(\gamma_{1}, \gamma_{2})= 
W_2(\mu_0(\gamma_1), \mu_0(\gamma_2)),
\quad 
   d_\mathcal{F}(\gamma_{1}, \gamma_{2})= 
W_2(\mu(\gamma_1), \mu(\gamma_2)),
\end{equation*}
Since $\mu(\gamma)$ has three components 
$(\sin\theta, \cos\theta, a)$, the squared distance 
$d_{\mathcal{F}}^2$ decomposes as:
\[
d_{\mathcal{F}}^2(\gamma_1,\gamma_2) = 
d_{\sin}^2 + d_{\cos}^2 + d_{a}^2,
\]
where $W_2$ is the vector-valued distance defined in Section \ref{stateofthe art}. Therefore, we propose the notion of connectivity kernel in the space of fragments:
\begin{equation}\label{KF}
K_\mathcal{F}(\gamma_i, \gamma_j) = e^{-\frac{d_{\sin}^2}{\sigma_1} + \frac{d_{\cos}^2}{\sigma_2} + \frac{d_{a}^2}{\sigma_3}};
\end{equation}
The positive constants $\sigma_i$ will be choosen in the sequel. An analogous notion of connectivity can be introduced for the distance 
$d_{0, \mathcal{F}}$. 

We suppose that this kernel models the cortical connectivity.
We do not claim that neurons explicitly compute an optimal 
transport plan. The Brenier approach shows that transport 
can be described in terms of geodesics in the space of 
probability measures, an approach recently extended to the 
sub-Riemannian setting. From a biological perspective, the 
activity measure $\mu_0(\gamma)$ defined in equation \ref{cifa} represents 
the probability that a kinematic feature is attained along 
the trajectory. We propose that neural activity evolves 
along paths constrained by anatomical connectivity encoded 
in the kernel $\omega_F(\gamma_i,\gamma_j)$. In this sense, 
transport along Wasserstein geodesics emerges naturally as 
a description of how activity propagates efficiently under 
these anatomical constraints, with cortical dynamics 
approximating such geodesic flows through local synaptic 
interactions and network structure. Optimal transport 
therefore provides an effective geometric description of 
neural population dynamics, rather than a directly 
implemented computational mechanism.







\subsection{Clustering Algorithm and generation of fragments}\label{sc}
 
\subsubsection{Brain activity}


In this section, we 
apply the spectral clustering algorithm with the proposed distance. The scope is to introduce an algorithm which can be implemented by the motor cortex. We make the assumption of a hierarchical brain organization in which the higher areas plan complex movements, and the fragments are organized in progressively shorter curves, while moving to the primary visual cortex. From this perspective, the clusterization of features in fragments (studied in \cite{mazzetti2023functional}) has to be compatible with the clusterization of fragments in neural states, to be studied here. 

Both are neurally based, and expressed in terms of the cortical activity equation proposed by  Amari \cite{amari1972characteristics} and Wilson and Cowan \cite{wilson1972excitatory}, and largely developed in the literature (see \cite{ermentrout1980large, bressloff2003functional, faye2010theoretical, sarti2015constitution}). In the space of fragments, the  connectivity kernel can be modelled as follows: 
\begin{equation}
\label{eq:meanField}
\frac{da(\gamma, t)}{dt}= -\nu a (\gamma, t)+ \mu\varrho\left(\int_{\Omega} K_{\mathcal{F}}(\gamma, \gamma')a(\gamma', t)d\gamma' + h(\gamma, t)\right),
\end{equation}
where $t > 0$, the coefficients  $\nu$ and $\mu$ represent the decay of activity and short-term synaptic facilitation, respectively. 
The function $\varrho$ is the activation function, typically a sigmoid or a ReLU, and $h$ is the input.\\ 
Following \cite{sarti2015constitution} and \cite{bolelli2025individuation}, we provide here 
the full stability argument. Since $K_F \in L^2(\Omega\times\Omega)$, 
the operator $A(a) = \mu\int_\Omega K_F(\gamma,\gamma')a(\gamma')d\gamma'$ 
is bounded and Lipschitz on $L^2(\Omega)$, guaranteeing existence and 
uniqueness of a solution to equation\ref{eq:meanField}.

For stability, consider the perturbation $u := a - a_0$ around a 
stationary state $a_0$. The linearized equation admits the Lyapunov 
functional $V(u) = \frac{1}{2}\int_\Omega u^2\,d\gamma$, whose 
time derivative satisfies:
\[
\frac{dV}{dt} \leq \left(-\nu + 
\mu\|K_F\|_{L^2(\Omega^2)}\right)\|u\|^2_{L^2}.
\]
Hence $a_0$ is asymptotically stable whenever 
$\|K_F\|_{L^2(\Omega^2)} < \nu/\mu$, which holds for 
physiologically plausible parameters. The dominant eigenfunctions 
of the eigenvalue problem
\[
\mu\int_\Omega K_F(\gamma,\gamma')\phi(\gamma')d\gamma' = 
\tilde{\lambda}\,\phi(\gamma)
\]
correspond to the stable emergent neural states. On the discrete 
fragment set $\{\gamma_i\}$, this reduces to $A_W\mathbf{u} = 
\tilde{\lambda}\mathbf{u}$, which is exactly the eigenvalue 
problem solved by spectral clustering on the affinity matrix 
$A_W$ (equation \ref{affinitymatrixwas}). Therefore spectral clustering is the direct 
numerical implementation of this stability analysis, not an 
external algorithm imposed on the data. Specifically, the stable neural states of equation~\ref{eq:meanField} correspond
to the dominant eigenvectors of the linearized connectivity kernel $K_\mathcal{F}$.
Spectral clustering extracts exactly these eigenvectors. Therefore, applying
spectral clustering to $K_\mathcal{F}$ is mathematically equivalent to finding stable 
emergent states of the cortical dynamics. It is a direct proxy for the
self-organizing process of M1, not merely an external algorithm imposed on the
data. Based on these arguments, we can find the emergent fragments by means
of spectral clustering.

\subsubsection{Implementation of the algorithm}

The choice of a distance, together with the definition  of connectivity kernel in the space of fragments given in \eqref{fragments}, leads to the matrix:
\[
\omega_\mathcal{F}(\gamma_i, \gamma_j) = e^{-d_\mathcal{F}(\gamma_i, \gamma_j)^2}.
\]

We postulate that this kernel can model the cortical connectivity, and we apply a spectral clustering algorithm to mimic the ability of the brain to cluster fragments into states.

 Let us summarize here the 
 main steps of the algorithm we propose:
The algorithm consists of two steps.

\textbf{Step 1 (Fragment generation):} Following 
\cite{mazzetti2023functional}, as recalled in Section~\ref{stateofthe art}, 
a spectral clustering algorithm is applied in the 
sub-Riemannian space $\mathcal{M}$ to organize the 
kinematic features $(x,y,\theta,t,v,a)$ into fragments 
$\gamma_i$, each a horizontal curve solution of 
equation~\ref{fragments}.
The connectivity kernel 
\eqref{KM}
is discretized in a discrete affinity matrix $ A_{\mathcal{M}}$

\textbf{Step 2 (Neural state generation):}
\begin{itemize}
\item Associate a density to each fragment $\gamma_i$ 
using $\mu_0(\gamma_i)$ (equation~\ref{cifa}) or 
$\mu(\gamma_i)$ (equation~\ref{density}).

\item Compute the Wasserstein distance 
$d_{\mathcal{F}}(\gamma_i,\gamma_j)$ using 
formula~\ref{Wass-d} for each component.
\item Compute the affinity matrix for every $(i,j)$ we define the   Wasserstein affinity matrix $A_{\mathcal{F}}$ in terms of the distances between the couple of curves $\gamma_i$ and $\gamma_j$ and the connectivity kernel \eqref{KF}. Precisely, 
\begin{equation}\label{affinitymatrixwas}
 A_{\mathcal{F}}(i, j) = \exp \left(-
 \frac{d_{\sin}^2(\gamma_i, \gamma_j)}{\sigma_1} - \frac{d_{\cos}^2(\gamma_i, \gamma_j)}{\sigma_2} - \frac{d_{a}^2(\gamma_i, \gamma_j)}{\sigma_3};
\right)   
\end{equation}
where \(\sigma_i\) are  parameters controlling the kernel width.

\item Apply spectral clustering to $A_W$ to obtain 
the neural states.
\end{itemize}


\section{Results}\label{results}
    
We generate a set of curves in 6-dimensional space. These curves are represented in two figures: the first Figure captures the position and orientation parameters $(x, y, \theta)$, while the second figure shows time, velocity and acceleration (t, v, a). 

\subsection{Clusterization with the real-valued probability: speed and curvature}
In this first example we  
consider a family of curves, in the 6D space $\mathcal{M}$, defined in \eqref{fragments}, and we apply our clusterization algorithm with the real-valued probability density introduced in definition 
\eqref{cifa}.

\begin{figure}[H]          \centering
\includegraphics[scale= 0.35]{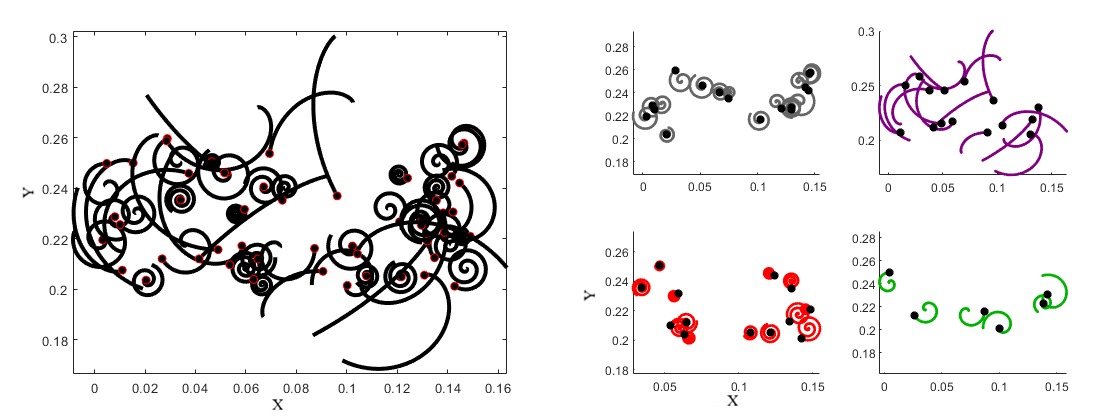}
\caption{A family of curves with different curvature (left) and its clustering with respect to curvature (right).}
\label{3.1fig}
\end{figure}

In order to understand the meaning of variables, we will start by considering  the projection of the curves in the space 
$(x,y,\theta)$. 
We consider 50 curves, solutions of the Cauchy problem \eqref{fragments}, with different values of $\alpha_2$, with $\alpha_3=0$, and random initial conditions.
It follows that $a$ is constant, and $v$ is linearly increasing or decreasing. 
Due to the relation \eqref{curv} between  $k$ and $v$, also the curvature is increasing or decreasing. Consequently, the curves are spirals (see Figure \ref{3.1fig}).  
To each curve, we associate the real value  density introduced in definition 
\eqref{cifa}.

Due to our choice of family of curves, and the definition of norm given in \cite{mazzetti2023functional}.
$$||\gamma'(t)|| = \sqrt{1 + \alpha_2^2}.$$
Hence the density value remains constant along the curves. It depends on the length of the lifted sub-Riemannian curve, which is expressed in terms of the length and the curvature of the 2D projection.

The spectral clustering algorithm introduced in subsection\ref{sc} is applied and produces four separate clusters (see Figure \ref{3.1fig}). 
Since the product of the length and the curvature is the number of times the curve twists on itself, the classification depends on this coefficient, which has a clear,  intrinsic topological meaning.

\subsection{Clusterization of artificially generated curves with the vector-valued probability}
In this second experiment, we test only the second step of the 
algorithm, which is the part introduced in this work. 
We therefore directly generate fragments $\gamma_i$ 
as solutions of equation~\ref{fragments}, bypassing Step~1 of the algorithm. Here, we apply a clusterization based on the vector-valued measure \eqref{density}
. 

\begin{figure}[H]
\centering
 \includegraphics[width=6cm,height= 4cm]
          {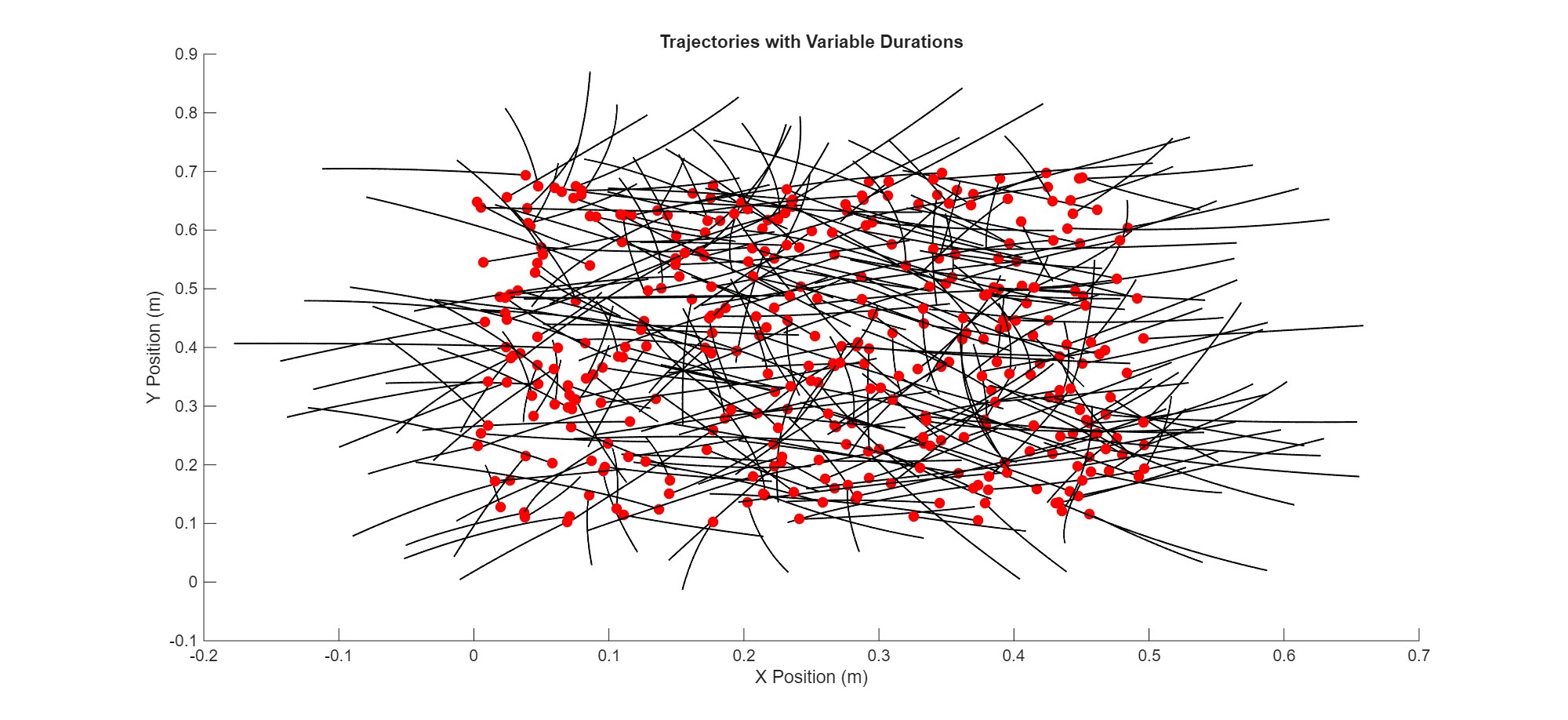}
          \includegraphics[width=6cm,height= 4cm]
          {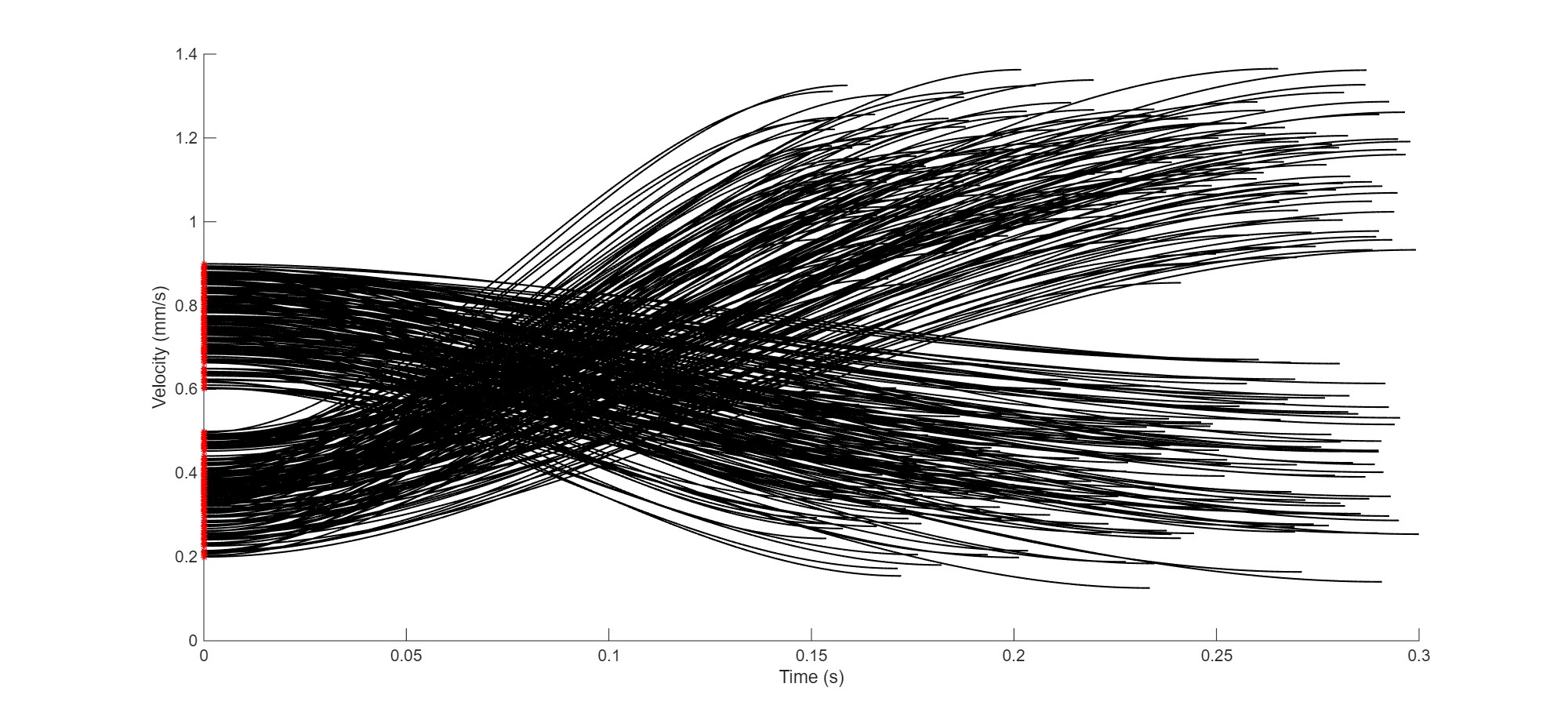}
\caption{Trajectories with random initial positions, orientations, and velocity profiles (accelerating or decelerating).}
\label{1155}
\end{figure}

A total of 350 trajectories were simulated to mimic biological motion, each consisting of 200 time points over a random duration of 0.15–0.3 s. Trajectories were generated with random initial positions, orientations, and velocity profiles (accelerating or decelerating) shown in the Figure \ref{1155}. Orientation changed linearly with a random angular velocity, and positions were computed by integrating velocity along the heading. Sample trajectories are illustrated in Figure \ref{1155} below, showing variability in shape, starting points, and orientation.

\textbf{Clustering based on Wasserstein distance}

\begin{figure}[H]
\centering
 \includegraphics[width=6cm,height= 4cm]{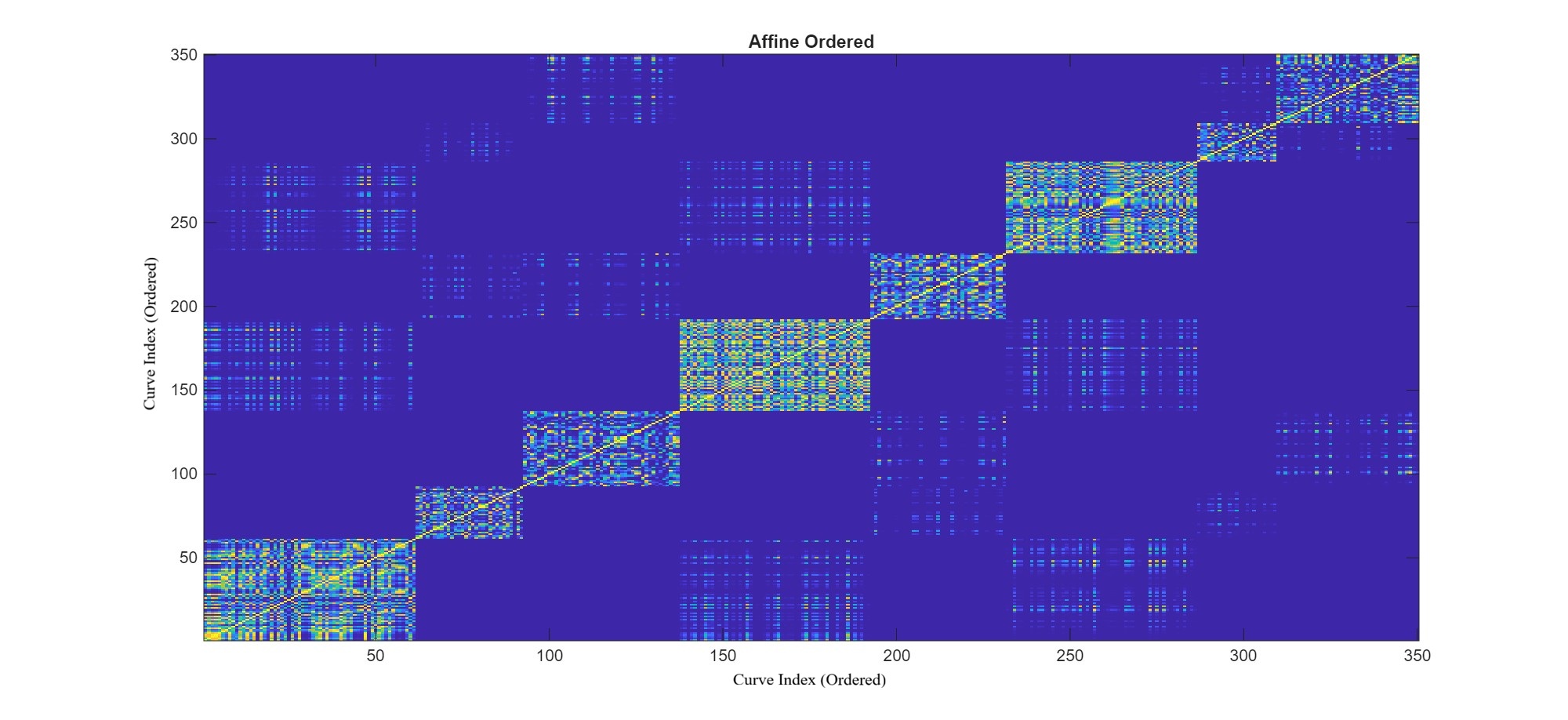}
\caption{Wasserstein affinity matrix.}
\label{1156}
\end{figure}

We apply the algorithm previously described and associated with vector valued probability measure \eqref{density}. We first computed Wasserstein distances between trajectories. The pairwise distances were converted into an affinity matrix and subjected to spectral clustering. The Wasserstein affinity matrix is shown in Figure \ref{1156}, highlighting strong block-diagonal structure.

\begin{figure}[H]
\centering
 \includegraphics[width=7cm,height= 4cm]
          {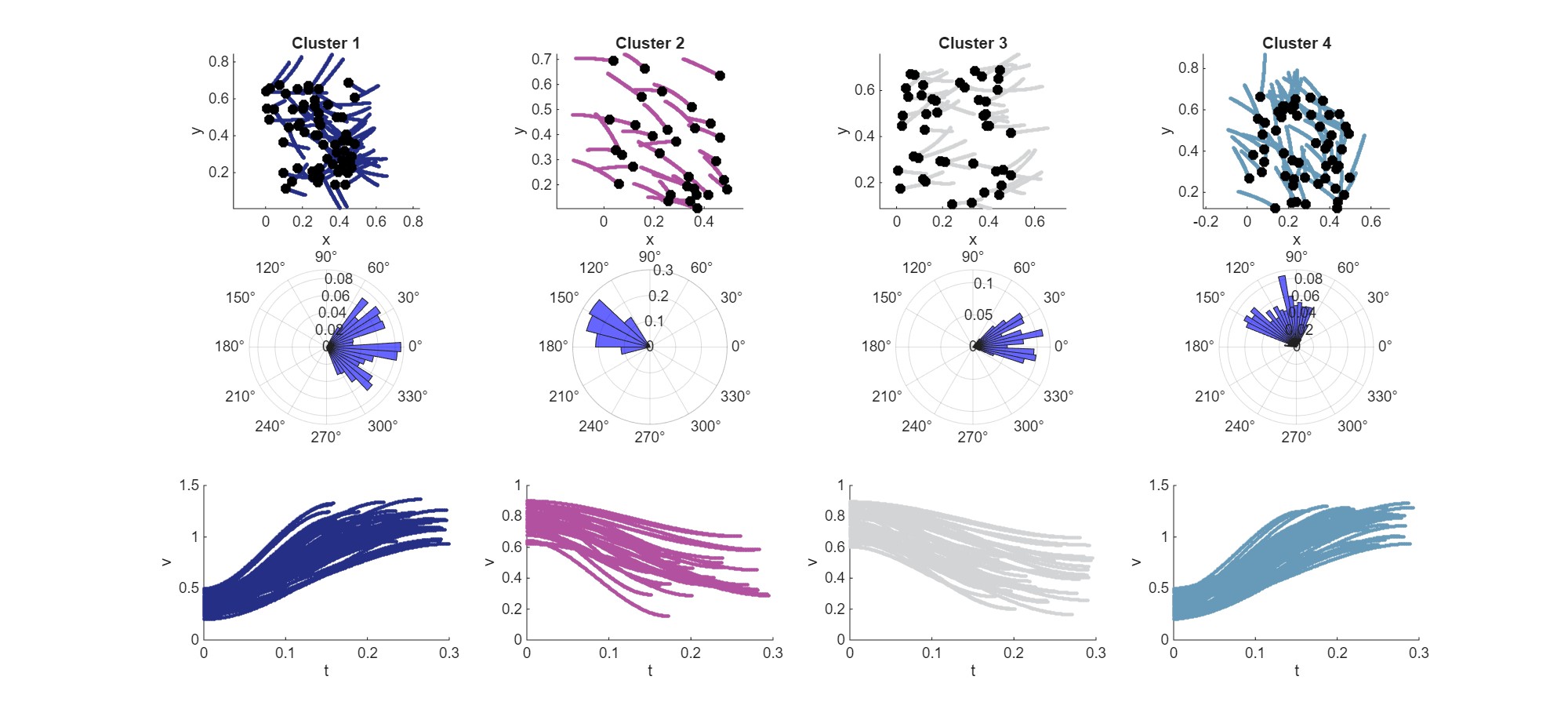}
          \hspace{ - 0.8cm}
          \includegraphics[width=7cm,height= 4cm]
          {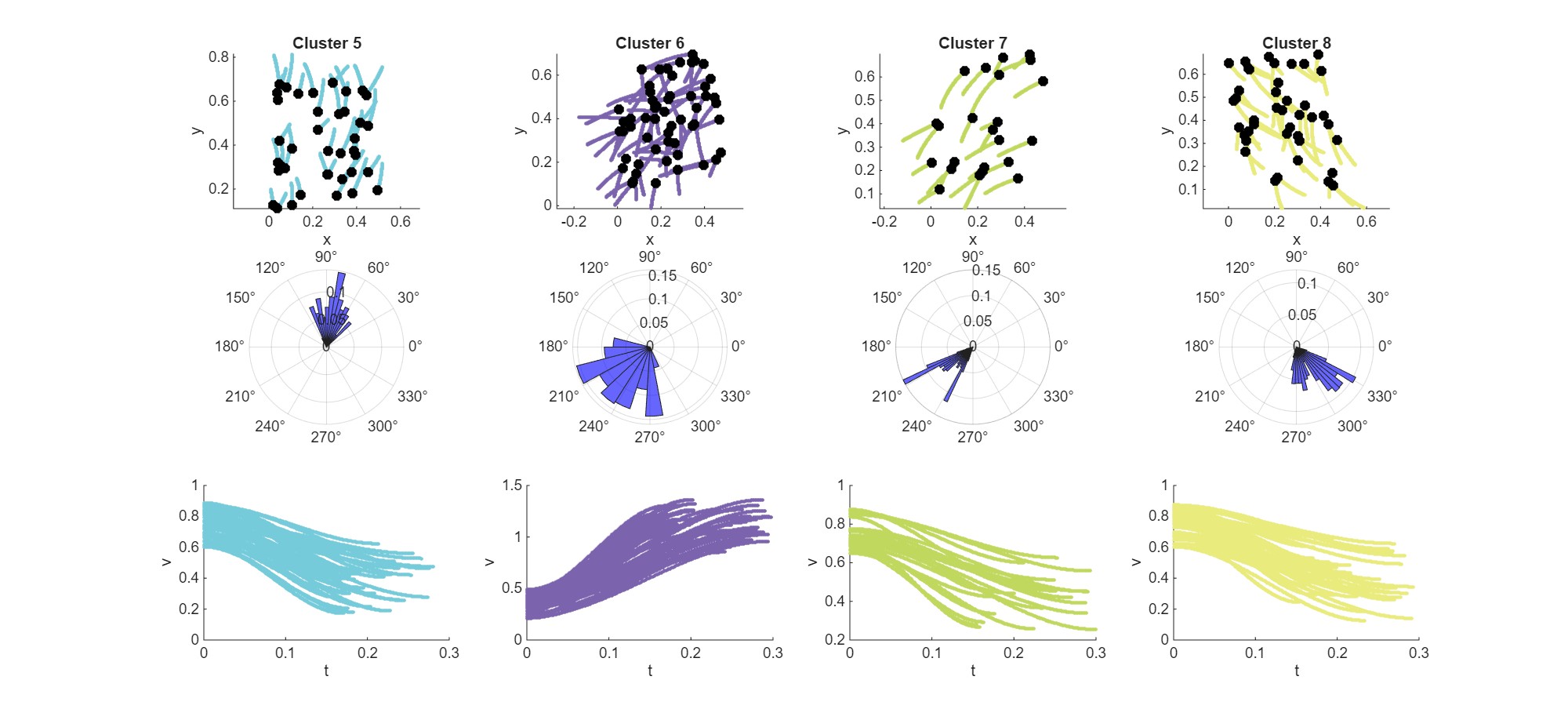}
\caption{Visualization of the grouping of fragments in neural states with Wasserstein distance.}
\label{1157}
\end{figure}

Eight clusters were identified, exhibiting high intra-cluster similarity. Silhouette score: 0.65, indicating well-separated clusters. 
In Figure \ref{1157}, trajectories colored by cluster, with starting points indicated, and demonstrate coherent grouping by trajectory shape and velocity profile.

\subsection{Clusterization of real movement trajectory fragments}\label{clusreal}
\begin{figure}[H]
\centering
 \includegraphics[width=6cm,height= 4cm]
          {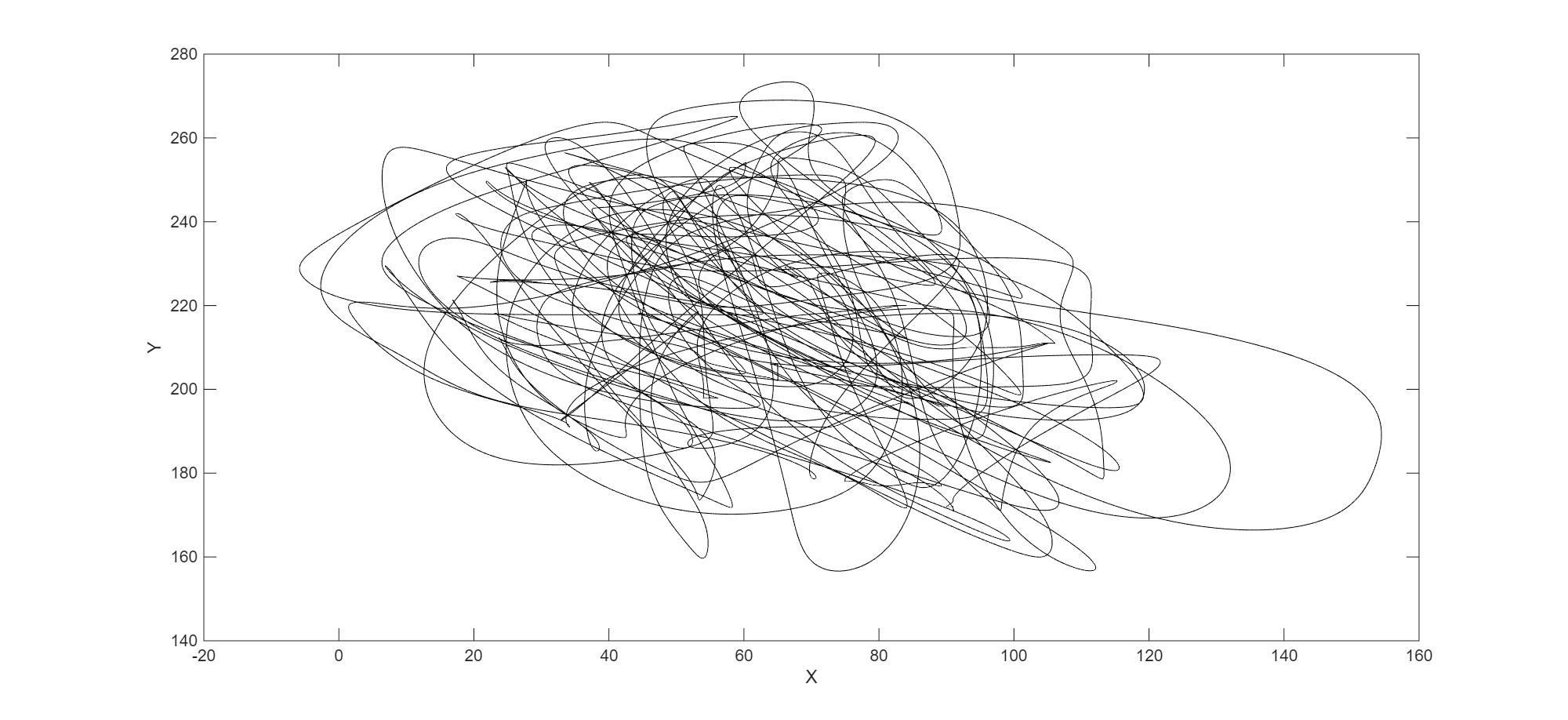}
          \includegraphics[width=6cm,height= 4cm]
          {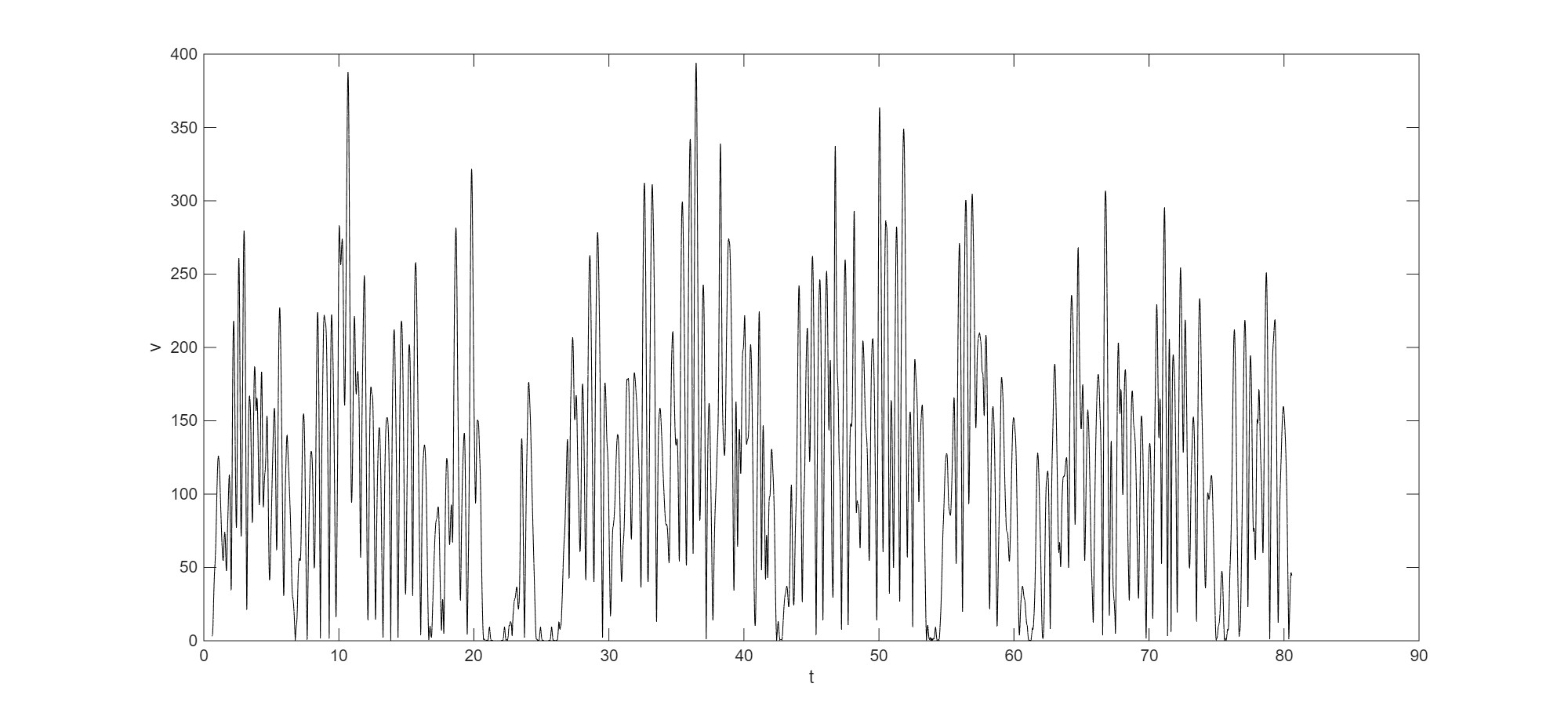}
\caption{Reaching path and speed profile of a center-out task.} 
\label{11133}
\end{figure}

In this third experiment, we apply the full two-step algorithm. 
We used the same data used in  \cite{kadmon2019movement} and kindly shared with us by Prof. Hatsoupolos. We extracted from the data set the \(x, y\) coordinates  of a total of 40,000 data points measured at fixed intervals of time \(t\) and 
we evaluated the corresponding values of 
\(v\) and \(a\). The input curve is shown in Figure \ref{11133}. 

{\bf Step one of the algorithm: clusterization into fragments.}
We start by applying the Sub-Riemannian classification algorithm proposed in \cite{mazzetti2023functional} to classify these features into distinct fragments.
In the context of the grouping problem, it is important to assign appropriate values to the constant coefficients $c_{i}$ in the exponential coordinates defined in equation \ref{distanza_periodic}. 
Since the temporal variable encodes velocity and acceleration information, we assign it a specific weight to reflect its role.  Note that the quotient of the distance over the parameter $\sigma$ is defined up to a constant which we can factor out. We normalize the time
$c_5 = 1$ (i.e., the time variable is not rescaled) 
since we consider this as an absolute independent variable
and $c_i = 1/0.008 = 125$ for $i = 1,\ldots,6$, 
with a rescaled kernel. The scope is to give a 
higher weight to 
the kinematic variables $(\theta, v, a)$ compared 
to the temporal increment, consistently with the 
observation that neural states are invariant to 
absolute temporal occurrence \cite{kadmon2019movement}. We also performed a sensitivity analysis by testing different weight configurations, and found that the chosen values consistently yield the most stable and robust clustering results (See Table \ref{table1} ).

\begin{table}[h!]
\centering
\caption{Sensitivity analysis of the coefficients \(c_i\) in \ref{subD}. 
Clustering stability was evaluated using the Silhouette score and the recovery 
of the 8 neural states reported in \cite{kadmon2019movement}.}
\label{table1}
\begin{tabular}{c c c}
\hline
\textbf{Coefficient configuration} & \textbf{Silhouette score} &  \\
\hline
$c_5 = 1$, \quad $c_i = 1/0.002$ for $i \neq 5$ & 0.21  \\
$c_5 = 1$, \quad $c_i = 1/0.005$ for $i \neq 5$ & 0.29 \\
$c_5 = 1$, \quad $c_i = 1/0.008$ for $i \neq 5$ & \textbf{0.35} &  \\
$c_5 = 1$, \quad $c_i = 1/0.010$ for $i \neq 5$ & 0.31 \\
$c_5 = 1$, \quad $c_i = 1/0.020$ for $i \neq 5$ & 0.24  \\
\hline
\end{tabular}
\end{table}

\begin{figure}[H]
\centering
 \includegraphics[width=6cm,height= 4cm]
          {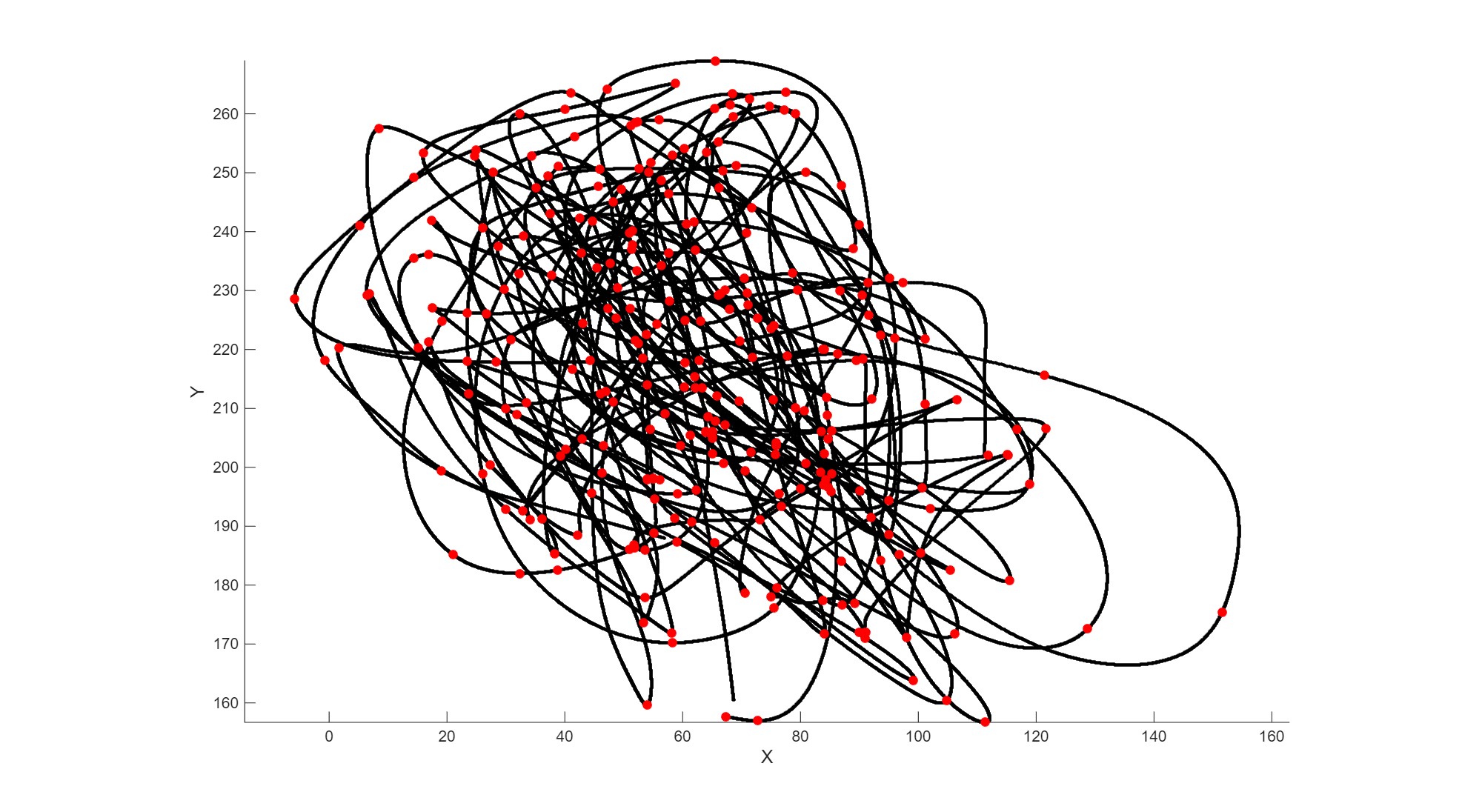}
          \includegraphics[width=6cm,height= 4cm]
          {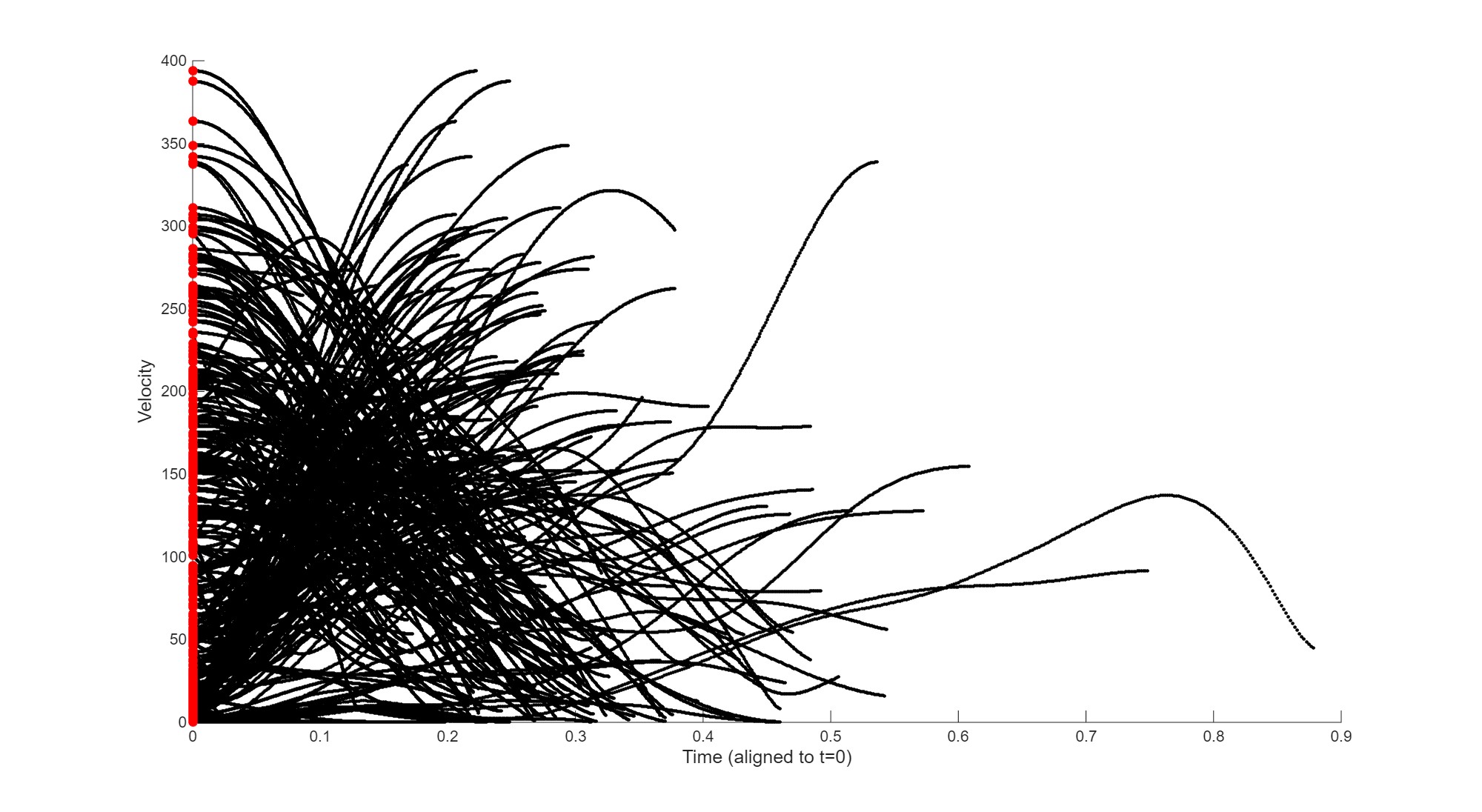}
\caption{Space of features decomposition into fragments. Each curve shows a single fragment.}
\label{1113}
\end{figure}

After applying the grouping algorithm, we obtained 355 initial fragments. To ensure meaningful analysis and correct classification into neural states, we filtered out clusters containing fewer than 10 data points. After this refinement, we obtained 329 fragments, as shown in Figure \ref{1113}.

We allow time shifting, but not dilation in the time variable. 
Biologically, this approach assumes that neurons possess a short-term memory buffer that retains a representation of the stimulus shape, enabling comparison of fragments that occur at different time instants. Such temporal alignment is consistent with the notion that certain neural populations encode shape features of movement trajectories independent of their absolute temporal occurrence, as supported by findings in the Kadmon paper\cite{kadmon2019movement}, where similar normalization is employed.

Consequently, all fragments are reparametrized 
with the same initial instant of time, even though these points correspond to different physical time instants.



 \textbf{Classification into neural states}\\

\begin{figure}[H]
\centering
\includegraphics[width=6cm,height= 4cm]{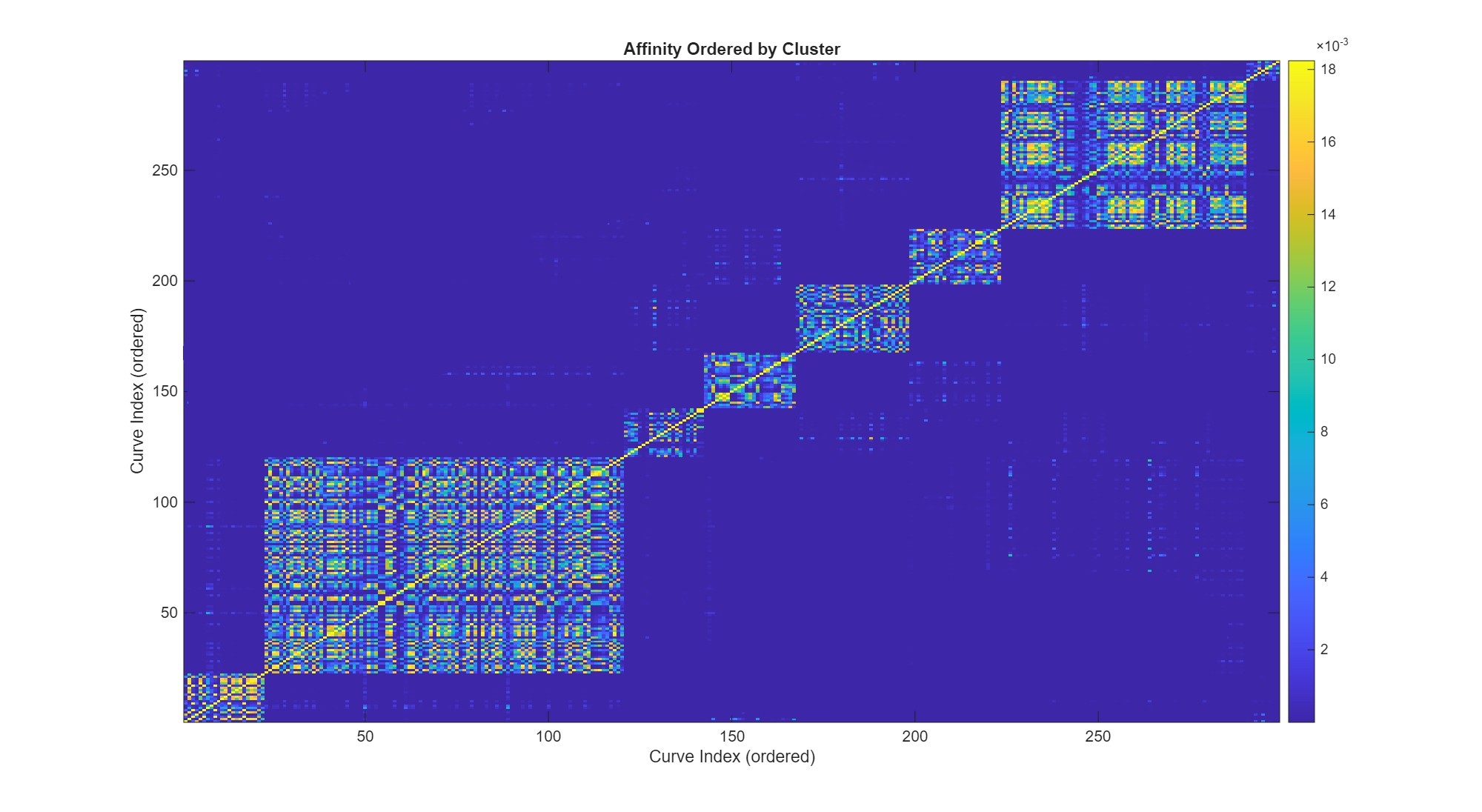}
\caption{Wasserstein affinity matrix. Indices have been re-ordered for visualization purposes.}
\label{WASSmat}
\end{figure}

\begin{figure}[H]
\centering
\includegraphics[width=7 cm]{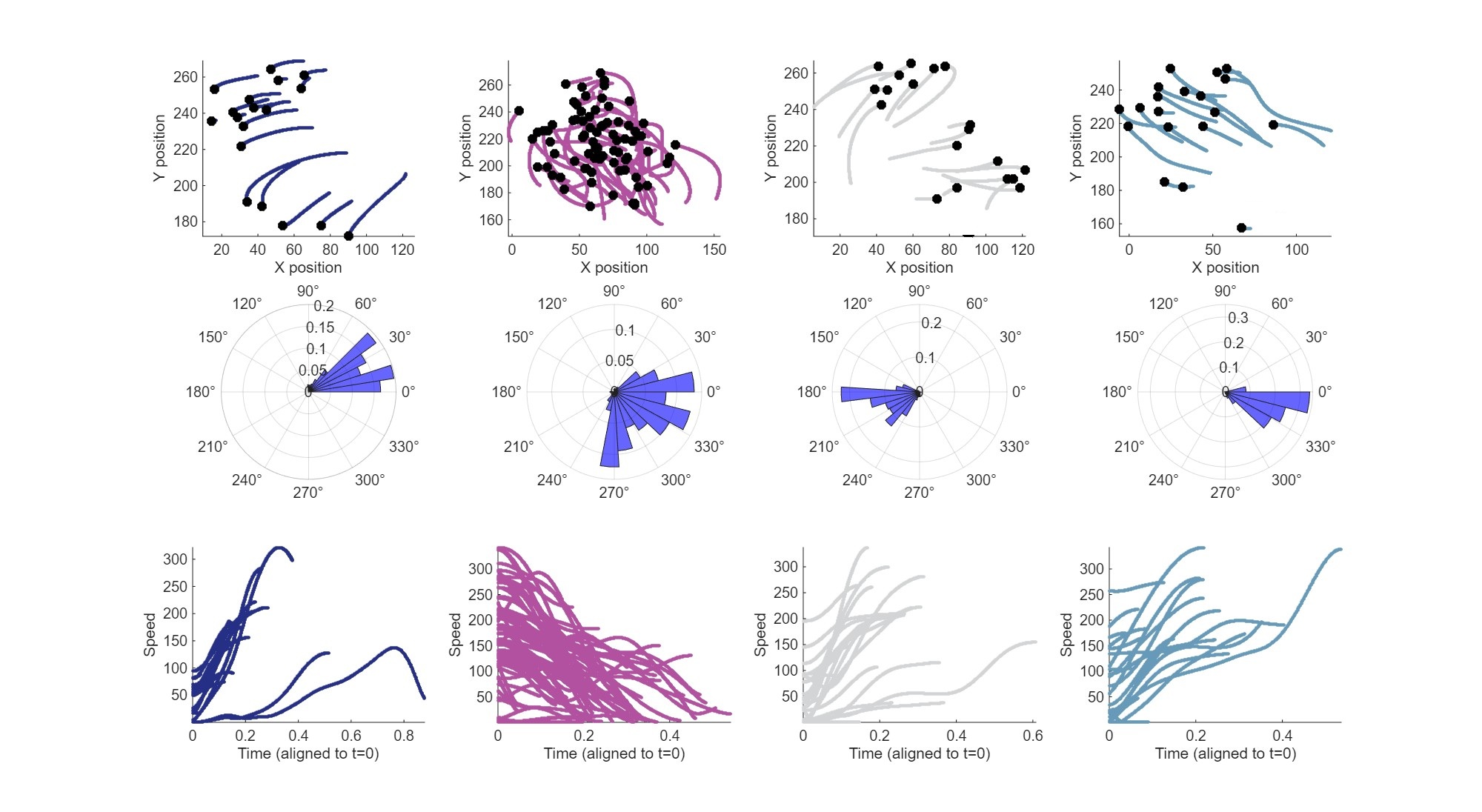}\hspace{ - .8cm}
\includegraphics[width=7 cm]{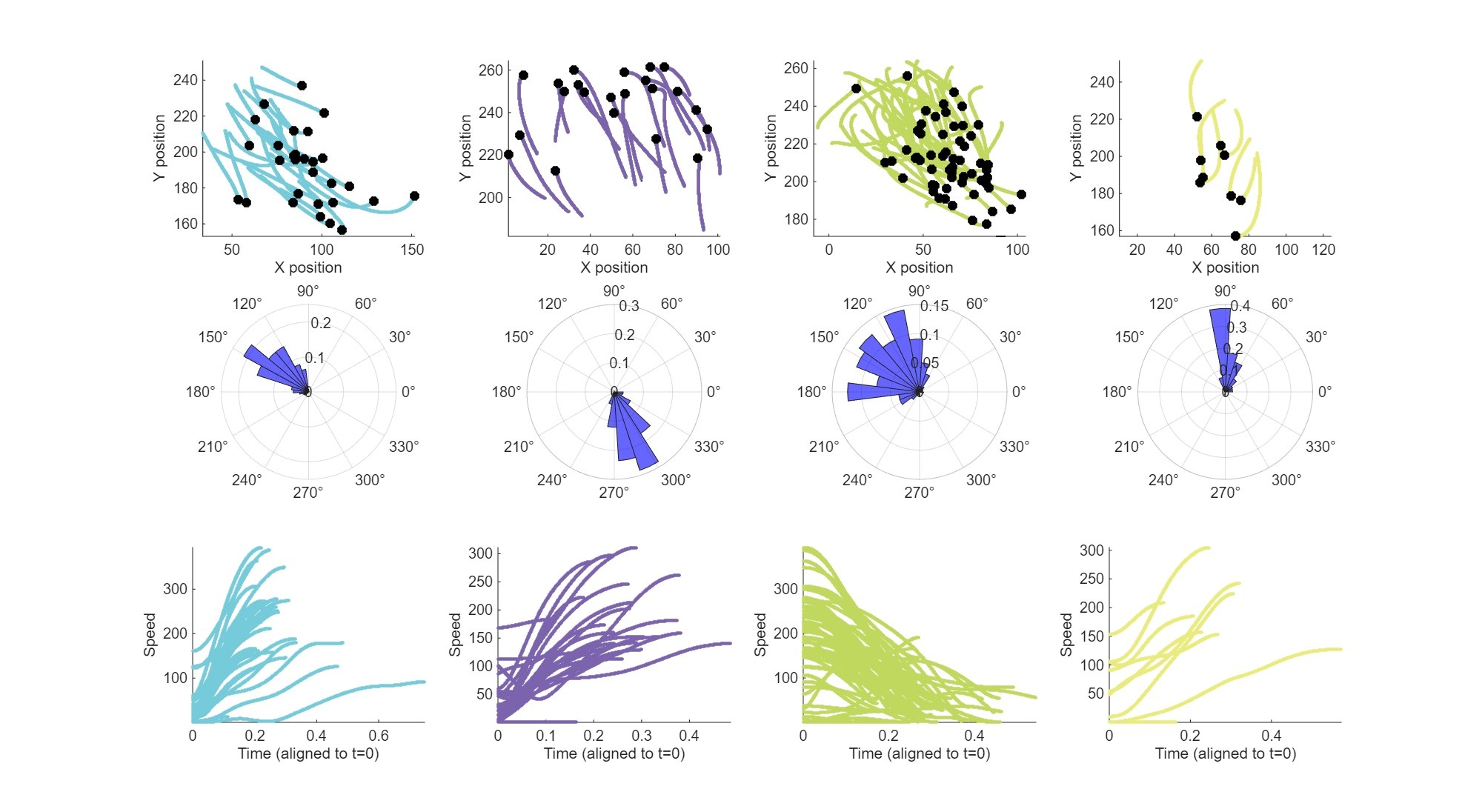}
\caption{Grouping of fragments in neural states with Wasserstein distance.}
\label{WAS}
\end{figure}

The next step is to apply the one-dimensional Wasserstein distance, as defined in equation~\eqref{Wass-d}, to categorize the fragments. The affinity matrix (Figure~\ref{WASSmat}) is computed using equation~\eqref{affinitymatrixwas} in this way, 
\begin{equation}
    A_W(i,j) \;=\; 
    \exp\!\left(-\frac{d_W^{(a)}(\gamma_i,\gamma_j)^2}{2\sigma_1^2}\right)
    \cdot
    \exp\!\left(-\frac{d_W^{(\cos\theta)}(\gamma_i,\gamma_j)^2}{2\sigma_2^2}\right)
    \cdot
    \exp\!\left(-\frac{d_W^{(\sin\theta)}(\gamma_i,\gamma_j)^2}{2\sigma_2^2}\right).
\end{equation}where we set the parameters $(\sigma_1 = 0.7)$ and $(\sigma_2 = 0.09)$ to enhance the sensitivity of the similarity measure to both orientation and acceleration features. Here $\sigma_1$ governs the kernel width for the acceleration feature, 
while $\sigma_2$ controls sensitivity for the orientation features
($\cos\theta$ and $\sin\theta$ share the same $\sigma_2$ by symmetry). 
The two bandwidths differ because the Wasserstein distances for acceleration 
and orientation live on different numerical scales: acceleration distances are 
substantially larger (hence $\sigma_1 = 0.7$), whereas orientation distances 
are much smaller (hence $\sigma_2 = 0.09$). 
Both values were set to approximately the median pairwise Wasserstein 
distance for their respective feature. A larger $\sigma$ broadens the kernel and densifies the affinity matrix, 
merging nearby trajectories into fewer, coarser clusters and generally 
increasing the Silhouette score at the cost of resolution. 
Conversely, a smaller $\sigma$ sharpens the kernel, producing finer-grained 
clusters but risking fragmentation and a reduced Silhouette score. 
To assess robustness, we swept $\sigma_1 \in [0.4,\,1.2]$ and 
$\sigma_2 \in [0.05,\,0.20]$: throughout this range the recovered cluster 
count remained stable at $8$ consistent with the neural states 
of \cite{kadmon2019movement}, and the Silhouette score varied between 
$0.58$--$0.65$ on artificial data and $0.22$--$0.38$ on real data. 
 The associated Silhouette score is 0.38, indicating well-separated clusters. Indeed, the Wasserstein distance performs effectively in distinguishing patterns among the fragments, as demonstrated in Figure~\ref{WAS}. We further performed a sensitivity analysis by varying the bandwidth parameters and observing the corresponding changes in clustering stability and Silhouette score, confirming the robustness of the selected configuration (see Table~\ref{table2}).
\begin{table}[h!]
\centering
\caption{Impact of different bandwidth parameters on clustering stability and Silhouette score.}
\label{table2}
\begin{tabular}{c c c c}
\hline
\textbf{$\sigma_1$} & \textbf{$\sigma_2$} & \textbf{Clusters} & \textbf{Silhouette Score} \\
\hline
0.4 & 0.05 & 9 & 0.22 \\
0.7 & 0.09 & 8 & \textbf{0.38} \\
1.0 & 0.12 & 8 & 0.33 \\
1.2 & 0.15 & 7 & 0.30 \\
\hline
\end{tabular}
\end{table}

\begin{figure}[H]
\centering
\includegraphics[width=7 cm]{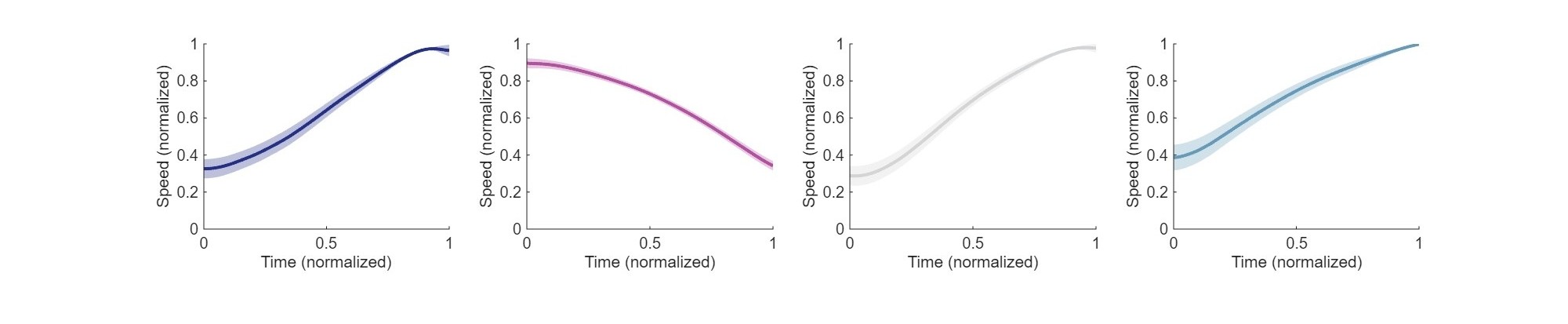}\hspace{ - .8cm}
\includegraphics[width=7 cm]{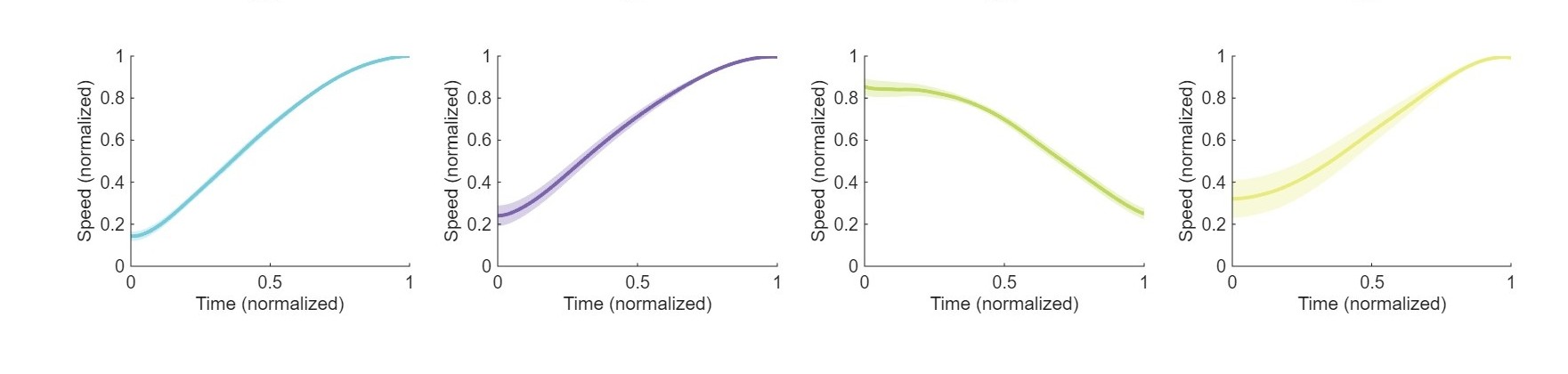}
\caption{Visualization of the grouping of fragments in neural states with Wasserstein distance.}
\label{visulization}
\end{figure}

For visualization and comparison purposes, the temporal domain of all fragments is normalized to the interval $[0,1]$ in the $(t,v)$ plane (see Figure~\ref{visulization}). This normalization allows comparison 
with Figure \ref{clusterini}, 
where the same normalization was applied. The clustering results obtained here are validated against the neural 
states of \cite{kadmon2019movement}, derived directly from 100-electrode array 
recordings in M1. Both our model and \cite{kadmon2019movement} produce 8 clusters 
that match in kinematic content: each cluster groups fragments with 
homogeneous orientation and a specific acceleration or deceleration 
profile (compare Figure \ref{visulization} with Figure \ref{clusterini}). The grouping in 
\cite{kadmon2019movement} was based on measured cortical activity, while ours 
is based purely on kinematics. The match between the two confirms that 
the sub-Riemannian geometry and the chosen kinematic variables are 
sufficient to explain cortical grouping, without requiring direct access 
to neural recordings. A direct neuron-by-neuron spike train correlation 
remains an important direction for future work.

\section{Conclusion}
In this study, we proposed a geometric model of the motor cortex that captures the hierarchical structure of motor primitives — from elementary features to complex neural states. A key contribution of our work is the demonstration that the well-known relation between movement speed and curvature — specifically, that curves with smaller radii are traversed more slowly than those with larger radii — naturally emerges from the geometry of fragments in our model. This supports experimental findings and further validates the relevance of using kinematic and geometric variables in modeling motor cortex activity.

We also evaluated our cortical grouping algorithm under the Wasserstein distance which provides more robust and reliable groupings. This is because the Wasserstein distance, is invariant with respect to time shift, whereas the Sobolev distance requires curves to be defined on the same domain, and compare speed and features at the same instant of time.

Finally, by applying our model-based grouping algorithm, we successfully recovered the same neural states as those identified by Kadmon-Harpaz et al \cite{kadmon2019movement}. from experimental cortical recordings. This confirms that our geometric modeling approach, the chosen variables, and the connectivity structure are sufficient to explain the organization of cortical activity, providing a principled and biologically plausible interpretation of how complex motor behaviors are encoded in the brain.

\bigskip

\textbf{Acknowledgments:} J.A. is funded by project MNESYS, PE12, PE0000006, 
G.C.. is funded by projects MNESYS, PE12, PE0000006, and by 
PRIN 2022 F4F2LH - CUP J53D23003760006

\bibliographystyle{ieeetr}
 
\bibliography{neurogeometry2}

\end{document}